\documentstyle[preprint,aps,epsf]{revtex}
\begin{document}
\def\be{\begin{equation}}
\def\ee{\end{equation}}
\def\bear{\begin{eqnarray}}
\def\eear{\end{eqnarray}}
\def\E{{\rm e}}
\def\bearst{\begin{eqnarray*}}
\def\eearst{\end{eqnarray*}}
\def\peleven{\parbox{11cm}}
\def\peffec{\peight{\bearst\eearst}\hfill\peleven}
\def\pspace{\peight{\bearst\eearst}\hfill}
\def\ptwelve{\parbox{12cm}}
\def\peight{\parbox{8mm}}
\title
{Gravitational instabilities and faster evolving density perturbations}
\author
{Elcio Abdalla and 
Roya Mohayaee}
\address
{\it Instituto de 
F\'\i sica-USP, C.P. 66.318, S\~ao Paulo\\
}
\date{01/10/98}
\maketitle

\begin{abstract}
\noindent

The evolution of inhomogeneities in a spherical collapse model is
studied by expanding the Einstein equation in powers of
inverse radial parameter.
In the linear regime, the density contrast is obtained
for flat, closed and open universes. 
In addition
to the usual modes, an infinite number of new growing modes are 
contained in the
solutions for pressureless open and closed universes.
In the nonlinear regime, we
obtain the leading growing modes in closed forms for a flat
universe and also, in the limits of small and large times,
for an open universe.

\end{abstract}
\hspace{.2in}


\section{Introduction}

The Newtonian and the relativistic 
theories of small perturbations form the 
theoretical basis for our present understanding of 
the formation of structures in the universe
\cite{weinberg,kolb,prad}.
In the Newtonian regime, which is appropriate for pressureless universes,
the hydrodynamic 
equations, incorporating the expansion of 
the universe, are perturbed by small fluctuations. 
Subsequently, a second-order differential 
equation is obtained whose solutions give the decaying and the growing
modes of the density contrast. This
quantity measures the evolution rate of the perturbation
with respect to the background density
and thus determines whether
structures can evolve from small initial perturbations or not. 

In the relativistic regime, one expands 
the Einstein metric $g_{\mu\nu}$ around the Friedmann-Robertson-Walker
metric by a local perturbation, $h_{\mu\nu}$, which arises naturally 
in an imperfect fluid model of the universe.  
Consequently, the evolution of the perturbations is 
formulated by means of the energy conservation equation, which 
can be written in terms of the time derivatives only, and a 
combination of the linearized Einstein equations 
(see {\it e.g.} \cite{weinberg}, chapter 
15). Space derivatives are eliminated in this procedure and a 
second-order ordinary 
differential equation in time is obtained for the density contrast.

The procedure in the relativistic regime can be summarized as
follows. Fixing the gauge such that the perturbation vanishes
in the ($i0$) and ($00$) directions,
one obtains, for the space-space component of the Einstein equation, 
\bear
&&\nabla^2 h_{ij}-\frac{\partial^2h_{ik}}{\partial x^j\partial x^k}-
\frac{\partial^2h_{jk}}{\partial x^i\partial x^k}+
\frac{\partial^2h_{kk}}{\partial x^i\partial x^j}-R^2\ddot h_{ij}\nonumber\\
&&+R\dot R\left(\dot h_{ij}-\delta_{ij}\dot h_{kk}\right)+2\dot R^2
\left( -2 h_{ij} +\delta_{ij}h_{kk}\right)=\label{w1}\\ 
&& -8\pi G \left( \rho -p\right) R^2 h_{ij} - 8\pi G R^4\delta_{ij} \left(
\delta\rho -\delta p\right) + \left( \hbox{ imperf. fl. correc.}\right),
\nonumber
\eear
where the last term corresponds to the imperfect fluid corrections
which are given in terms of a velocity field.
The differential equation contains space derivatives of the
metric and, therefore, is extremely difficult to solve. 
In general, matter distribution constrained to a small
region has large space derivative as compared to the metric itself.
This is clear for a well-localized Gaussian distribution, 
$\rho(\vec x)\sim {\rm e}^{-\lambda (\vec x-\vec x_0)^2}$,
for large $\lambda$, which cannot be
expanded in inverse powers of the distance $\frac 1{\vert\vec x-\vec x_0
\vert}$. 
The off-diagonal components of the Einstein equations are not amenable 
either, once again, due to the space derivatives. 
However, the time-time component
contains no space derivative and reads

\be
\ddot h_{kk} -2\frac{\dot R}R \dot h_{kk} +2\left( \left(\frac{\dot R}R\right)^2
-\frac{\ddot R}R\right) h_{kk}=-8\pi G \left(\delta\rho +3\delta p\right)R^2.
\label{hh}
\ee
Additional informations can also be obtained from the energy-momentum
conservation equation \footnote{It is worth commenting that 
the inclusion of higher derivatives may introduce zero
modes.},
\bear
\dot{\delta\rho} &+& 3\frac{\dot R}R \left( \delta\rho -\delta p\right) =
\left(\rho - p\right)  \frac{\partial}{\partial t}
\left(\frac{h_{kk}}{2r^2}\right) + \left(\hbox{imperf. fl. correc.}\right)
\label{dotdeltarho}  
\eear
which, together with a few general hypothesis,
leads to the simple relation $\delta(t) \equiv
\delta\rho/(\rho +p)=-h_{kk}/2R^2$. 
This relation, when used 
in (\ref{hh}), gives rise to a second-order 
differential equation
whose solutions give the growing and the decaying modes of the density
contrast $\delta(t)$.

In the nonlinear regime, 
evolution of mass density fluctuations has been studied in
the matter-dominated Einstein-de-Sitter universe 
by means of the power spectrum,
which is obtained from the Fourier transform of the density contrast
\cite{frieman}.

Although the standard procedure is formulated in such a way as to
enable us 
to describe the evolution of
a general kind of perturbation,
the hypothesis of an imperfect fluid with a velocity field which is at
the root of this scheme cannot be
applied to certain kinds of fluctuations. 
In the standard scheme,
the fluctuation seeds, which are believed to give origin to the
galaxies, are described by Poisson-like
perturbations. 
These are well-localized functions with large space derivatives.

Contrary to the standard scheme, we study 
soft perturbations, which
are almost homogeneous at large scales.
These perturbations are especially interesting for the purpose of studying
inhomogeneities at very large scales \cite{peebles,cosmopert}. 
They are also important for studying totally inhomogeneous universes 
\cite{inho-old,krasinski,inho-new,kras2} .
Although perturbative expansion
around a comoving metric is a subtle procedure, 
smooth perturbations at large scales 
can be incorporated into perturbative schemes. 
 
We consider the ho\-mo\-ge\-ne\-ous Fried\-mann\--\-Robert\-son-Wal\-ker 
metric as a background metric and expand 
the perturbation in powers of 
the inverse radial parameter, {\it i.e.} in powers of $1/r$. The 
advantage of this procedure, which is valid for 
very large scales, is that we can control the 
space-dependence of the perturbation at each order, 
and, as we shall see, the Einstein equations 
will provide us with more informations on the evolution of 
the perturbations.

In this approach, equation (\ref{w1})  simplifies
and all the space derivatives
disappear in the first
approximation (the zero-th order terms, being homogeneous, are zero
modes of space derivatives). In particular, 
even though (\ref{dotdeltarho}) is still valid,
we no longer
need higher-order time derivatives to determine the
density contrast. 

Our aim is to use the aforementioned expansion in
$1/r$ for spherically symmetric configurations,
{\it i.e.}, for the S-waves, and verify the presence of the growing modes for
the density contrast. 

We parameterize the metric as
\be
ds^2 = dt^2 -R_p(t,r)^2\left(\frac{dr^2}{1-kr^2} +r^2d\Omega^2\right)
\label{metric} 
\ee
where the scale factor $R_p(t,r)$ is a function of the radial parameter as
well as time. Using this metric in the Einstein equation we obtain
\bear
2\frac {\ddot {R_p}}{R_p} +&&\left( 1+3\omega\right) 
\left(\frac{\dot{R_p}}{R_p}\right)^2+
2kr\omega\frac{{R_p}^\prime}{{R_p}^3}
-\frac{{(1+3\omega)k}}{{R_p}^2}
\nonumber\\
-&&\left(\frac 2r\left( 1+2\omega\right)\frac {{R_p}^\prime}
{{R_p}^3}+\left( 1-\omega\right)\frac {{{R_p}^\prime}^2}{{R_p}^4}
+2\omega\frac {{{R_p}^{\prime\prime}}}{{R_p}^3}\right)
\left( 1-kr^2\right) 
=0.
\label{r-equation} 
\eear
where ``dot'' and ``prime'' stand for time and space derivatives, 
$k=0,1,-1$ for flat, closed and open
universes respectively and the parameter $\omega$ is the pressure-density 
proportionality constant, {\it i.e.} $P=\omega\rho$. 
The density is given by the time-time component of 
the Einstein equation
\be
3\left(\frac{\dot R_p}{R_p}\right)^2
+\left(-\frac 4r\frac {{{R_p}^\prime}}{{R_p}^3}
+\frac {{{R_p}^\prime}^2}{{R_p}^4} 
-2\frac {{{R_p}^{\prime\prime}}}{{R_p}^3}\right)\left( 1-kr^2\right)
+\frac {3k}{{R_p}^2}
+2kr\frac{{R_p}^{\prime\prime}}{{R_p}^3} 
=8\pi G\rho_p(r,t).
\label{rho-ee} 
\ee

To solve the above equations we expand the scale factor and the
density around the Friedmann background by small fluctuations in
powers of inverse radial parameter. That is,
\bear
R_p(t,r) &=& R+ \sum_{n=0}^\infty {\delta R}_n(t)r^{-n}\quad ,
\label{r-exp}\\ 
\rho_p (t,r) &=&\rho+ 
\sum_{n=0}^\infty {\delta\rho}_n(t)r^{-n}\quad ,\label{rho-exp} 
\eear
where $R$ and $\rho$ are the Friedmann scale factor and density 
respectively. It is worth commenting that these solutions are not
just gauge modes of the metric but represent genuine perturbations
 \footnote{For a comprehensive study of the gauge freedom arising 
in the context of density perturbations, we refer
the reader to reference \cite{ellis}.}.
It is easy to see that a solution $\zeta_\mu$ of $
\delta g_{\mu\nu}=\zeta_{\mu,\nu}+\zeta_{\nu,\mu}$ where $\delta
g_{\mu\nu}$ is the metric perturbation, such that
$\delta g_{0r}=0\,\, {\rm and}\,\, 
\delta g_{rr}\sim  t^a r^{-n}$, cannot be obtained.

In the following sections, we solve the perturbed
Einstein equations for different values of $k$ and $\omega$ and
find the density contrast which is defined as:

\be
\delta (t,r) =\frac { \delta\rho (t,r)}{\rho  (t)} \label{contrast} 
\ee
where $\delta\rho$ is given by the 
summation term in equation (\ref{rho-exp}).

\section{Matter-dominated open universe}

The scale factor in the Einstein equation
(\ref{r-equation}) is expanded as
in (\ref{r-exp}) and the coefficients at 
the $r^{-N}$ order are collected. For $\omega=0\,\, {\rm and}\,\,
k=-1$, these coefficients satisfy the recursive relation 
\bear
\sum_{n,m,l}&&\left( 2\ddot{{\delta R}_n}{\delta R}_m{\delta R}_l
{\delta R}_{N-n-m-l} + {\delta R}_n{\delta R}_m{\delta R}_l
{\delta R}_{N-n-m-l}\right) +\nonumber\\
&&+\sum_n n(4-N+n){\delta R}_n{\delta R}_{N-2-n}=0, 
\label{recurs-open-md} 
\eear
where ${\delta R}_n$ represents $R+{\delta R}_0$ for $n=0$ and the
indices $n,\,\,m\,\, {\rm and}\,\, l$ run from zero to infinity.
The above equation can be solved recursively for each $N$.
In general, using the parametric expressions 
$R=\cosh\psi-1$  and  $t=\sinh\psi-\psi$, we obtain
an inhomogeneous second-order 
differential equation, whose non-homogeneous terms are given in 
terms of ${\delta R}_{j<n}(\psi)$. Although, it is not possible 
to solve such an
equation in the general case, the homogeneous part of 
equation (\ref{recurs-open-md}),
\be
\frac{d^2{\delta R}_n}{d\psi^2} +\left( n-\frac 1{2\sinh^2\frac{\psi}2}
\right){\delta R}_n =0\quad , \label{diff-eq-open} 
\ee
can be solved recursively for each $n$.
The corresponding solutions are
\bear
({\delta R}_n)_+ &\sim&  \frac 1{\sqrt n \sinh(\psi/2)} \left(
\cosh(\psi/2)\sin\sqrt n\psi -2\sqrt n\sinh(\psi/2)\cos\sqrt n\psi\right)\quad ,
\nonumber\\
({\delta R}_n)_- &\sim& \frac 1{\sinh(\psi/2)}\left(\cosh(\psi/2) 
\cos\sqrt n\psi
+2\sqrt n \sinh(\psi/2) \sin\sqrt n\psi \right) \quad ,
\label{diff-eq-open-sol} 
\eear

Using this result in (\ref{rho-ee}), we obtain the 
density and subsequently the density contrast by using 
expression (\ref{contrast}). This is given by
the growing mode
\be
\delta_+(\psi)\sim\frac{\sin\left(\sqrt n\psi\right)}{\sqrt n}\left[
\frac{3\sinh\psi}{\left( 1-\cosh\psi\right)^2} +\frac {2n\sinh\psi}
{1-\cosh\psi}\right] +\cos\left(\sqrt n\psi\right)
\frac{5+\cosh\psi}{1-\cosh\psi} \quad ,
\label{delta-plus-open} 
\ee
and the decaying mode
\be
\delta_-(\psi)\sim\cos\left(\sqrt n\psi \right)\left[
\frac{3\sinh\psi}{\left( 1-\cosh\psi\right)^2} +\frac {2n\sinh\psi}
{1-\cosh\psi}\right] - \sqrt n \sin\left(\sqrt n\psi\right)
\frac{5+\cosh\psi}{1-\cosh\psi}\quad . \label{delta-minus-open} 
\ee
The $n=0$ ({\it i.e.} the $r$-independent perturbation) 
expressions reproduce the
standard growing mode
\be
\delta_+^{n=0}(\psi)\sim\frac{3\psi\sinh\psi}{\left( 1-\cosh\psi\right)^2} 
+\frac{5+\cosh\psi}{1-\cosh\psi}  \label{delta-plus-open-0}\quad ,
\ee
and the standard decaying mode
\be
\delta_-^{n=0}(\psi)\sim\frac{3\sinh\psi}{\left( 1-\cosh\psi\right)^2}
\label{delta-minus-open-0}\quad  
\ee
for the density contrast \cite{weinberg,kolb}.
The solutions (\ref{delta-plus-open}) and (\ref{delta-minus-open}) 
show that in addition to the standard growing mode, 
infinitely many growing modes exist at all higher orders in 
the perturbation expansion in the linear regime.

All the above results can be used for the closed
universe by making the substitution $\psi\equiv -i\theta$.

Non-linear growing modes would be expected if one could solve the
inhomogeneous differential equation (\ref{recurs-open-md}). 
In Section V, we
obtain these modes in the two extreme limits of small and large times.
In the small-time limit, open and close universes behave as a flat universe. 
In the asymptotic, {\it i.e.} large time, limit 
the scale factor 
varies linearly with time and the
inhomogeneous equation (\ref{recurs-open-md}) can once again 
be solved at higher orders. 
We obtain a new growing mode in this limit.

\section{ Matter-dominated flat universe}

For the flat universe,  the series expansion (\ref{r-exp})
of the Einstein equation (\ref{r-equation}) leads to the recursive equation 
\bear
&&\sum_{n,m,l,p}\left( 2\ddot {\delta R}_n {\delta R}_m {\delta R}_l 
{\delta R}_p +\dot {\delta R}_n\dot {\delta R}_m {\delta R}_l {\delta R}_p
\right) r^{-(n+m+l+p)} \nonumber\\
&&+\sum_{n,m}\left(n(2-m) {\delta R}_n {\delta R}_m \right) r^{-(n+m+2)}=0 
\quad  \label{einstein-expanded}
\eear
where the indices $n,m,l$ and $p$ run from zero to infinity.
Unlike, the differential equation (\ref{recurs-open-md})
for the open universe which could not be solved without discarding
the inhomogeneous terms, the
above inhomogeneous equation for the flat universe can be solved recursively 
to find the perturbation coefficients $R[n](t)$ 
\footnote{ The results have been
reproduced by a MapleV-5, program \cite{maple}.}. 

The first few coefficients are
\bear
{\delta R}_0 &=&  c_0 t^{\frac 23} \quad ,\\
{\delta R}_1 &=&  c_1 t^{\frac 23} + c_2 t^{-\frac 13} \quad ,\\
{\delta R}_2 &=&  c_1 t^{\frac 23} + c_3 t^{-\frac 13}-\frac 14 c_2^2 
t^{-\frac 13}\quad , \\
{\delta R}_3 &=& -\frac 9{10}c_1t^{\frac 43}+ c_4 t^{\frac 23} +\frac 92 c_2 
t^{\frac 13} \nonumber\\
&&+ c_5 t^{-\frac 13} + \left(\frac 14 c_1 c_2^2-\frac 12 c_2 c_3\right)
t^{-\frac 43}+\frac 16 c_2^3t^{-\frac 73}\quad ,\label{deltaR3}\\
{\delta R}_4 &=& \left(\frac 94 c_1^2 -\frac 95 \right) t^{\frac 43} 
+c_4 t^{\frac 23} 
+\nonumber \\
&&\left(-\frac {99}5c_1+9c_3\right) t^{\frac 13} + c_6 t^{-\frac 13}
+\frac{27}8c_2^2 t^{-\frac 23}\nonumber\\
&&+C_{dd}t^{-\frac 43}+C_at^{-\frac 73}+C_bt^{-\frac {10}3}\quad ,\\
{\delta R}_5 &=& \frac{243}{280}c_1t^2+\left( -\frac{27}{10}c_4 -\frac{81}{20}
c_1^3 +\frac{36}5c_1^2\right)t^{\frac 43}-\frac{243}8 c_2t +  c_7 t^{\frac 23}+
\nonumber\\
&&\left( -\frac {153}5 c_1c_2 -\frac{333}{10}c_1c_3 +\frac {27}2c_5+
\frac {513}{10} c_1^2c_2\right)  t^{\frac 13} +c_8 t^{-\frac 13}+\nonumber\\
&&\left( -\frac {693}{40} c_1c_2^2 +\frac{45}4c_2c_3\right)t^{-\frac 23} +
C_ct^{-\frac 43}+C_dt^{-\frac 73}+C_ft^{-\frac {10}3}\quad ,
\eear
where $C$'s are functions of either preceding $c$'s 
or are new constants themselves.
In general, the leading growing modes can be written in a closed form as
\be 
{\delta R}_+(r,t)=\sum_{n=0}^\infty c_{2n+3}^\prime
\frac {t^{2(n+2)/3}}{r^{2n+3}}\quad .
\ee

From the time-time component of the Einstein equation (\ref{rho-ee}), we 
recover the mass 
distribution and subsequently the density contrast 
$\delta (t,r) = \sum_{n=o}^\infty \delta_n(t)r^{-n}$, whose 
first few terms, up to an overall constant factor of 3/4, are
\bear
\delta_1 &=&  -4c_2 t^{-1} \quad , \\
\delta_2 &=&  4\left(c_1c_2-c_3\right) t^{-1}+ 9c_2^2t^{-2} \quad , \\
\delta_3 &=&  -\frac{12}5c_1t^{\frac 23} -6c_2 t^{-\frac 13}\nonumber\\
&&+4\left( c_1c_3+ c_1c_2+  c_1^2c_2-c_5\right) t^{-1} + 18\left( c_2c_3
- c_1c_3^2\right) t^{-2}-18c_2^3 t^{-3} \quad ,\\
\delta_4 &=& \left(-\frac {14}5 c_1  +\frac {47}5c_1^2 \right) t^{\frac 23}
+\left( \frac{184}5c_1c_2 - 16c_3   \right) t^{-\frac 13} + 
17 c_2^2t^{-\frac 43} \nonumber\\
&& 4\left(c_2c_4 +c_1c_5+c_1^3c_2 -c_6+c_1c_3-2c_1^2c_2-c_1^2c_3\right) 
t^{-1}+\nonumber\\
&&9\left( -4c_1c_2c_3 +2 c_2c_5 -2 c_1c_2^2 +3 c_1^2c_2^2 + c_3^3
\right)  t^{-2} + 54\left( c_1c_2^3 -c_2^2c_3\right) t^{-3}
+\frac{135}4c_2^4 t^{-4} \quad ,\\
\delta_5 &=&\frac{162}5 c_1 t^{\frac 43} +\left(-54c_2-\frac{81}2c_2
\right)t^{\frac 13}\nonumber\\
&&+\left( -12c_4-4c_1^3+16c_1^2+\frac{132}5c_1^2-\frac{96}5c_1^3-\frac{36}5c_4
\right) t^{\frac 23}+ C_{gg}t^{-\frac 13} + C_{kk} t^{-\frac 43}+ 
C_h t^{-\frac 73} \nonumber\\
&& C_i t^{-1}+ C_j t^{-2}+ C_k t^{-3}+  C_lt^{-4}+ C_m t^{-5} \quad.
\eear
In general, the leading growing modes 
can be represented in the closed form
\be
\delta_+=\sum_{n=0}^\infty c_{2n+3}^{\prime\prime}
\frac {t^{2(n+1)/3}}{r^{2n+3}}\quad.\label{flatdensitycontrast}
\ee

These results show that up to the third order
in the perturbation expansion, the density contrast does not grow. 
This is due to the fact, mentioned in the
introduction, that the perturbation coefficients at 
these orders are not sensitive to 
the space-dependence of the perturbed scale factor $R_p(t,r)$. However, 
the third-order perturbation term
contains the usual growing mode in the linear regime \footnote{
It is worth making the comparison with the open universe where
the standard zero modes already appear at the zero-th order.},
 namely $\delta\sim t^{2/3}$ . 
(This term emerges because of the $t^{4/3}$ growing mode in (\ref{deltaR3}).)
The nonlinear modes are given by $n\rangle 0$ terms in expression
(\ref{flatdensitycontrast}) and 
fully agree with the results of
reference \cite{frieman}.

Once again, we see that in addition to the usual growing modes
arising in the linear regime,
many more growing modes appear at higher 
orders in the perturbation series.
The higher growing modes can have interesting consequences 
for structure formation since 
different perturbative regimes can correspond to different orders in
the perturbation series.
For very large structures, such as superclusters, the dominating terms
are the first few, which vary with a smaller powers of time and so
are expected to appear later. On the other hand, small fluctuations
are described by higher-order terms in the perturbative series and
thus can form earlier. This simple argument points toward a bottom-up
scenario for the formation of structures and thus a
universe dominated by Cold dark matter \cite{zeldovich}.

\section{ Radiation-dominated flat universes}

In this case, we solve, using a MapleV-5 program \cite{maple}, 
a recursive equation similar to (\ref{einstein-expanded}) 
for the coefficients of the
scale factor. 
As in the previous
cases, the leading growing modes 
can be written in the following closed form:
\be 
{\delta R}_+(r,t)=\sum_{n=0}^\infty c_{2n+3}
\left(\frac {\sqrt t}{r}\right)^{2n+3}\quad .
\ee
Substituting this solution in the time-time component of the Einstein
equation (\ref{rho-ee}) and 
subsequently using (\ref{contrast}) we obtain the expression
\be 
\delta_+(r,t)=\sum_{n=0}^\infty {c^{\prime\prime}}_{2n+3}
\frac {t^{n+1}}{r^{2n+3}}
\quad 
\ee
for the leading growing modes of the density contrast.

Once again, in addition to the standard growing mode, $\delta\sim t$, which
arises at the third order of the series, we obtain an infinite
number of higher-order growing modes.

\section{Nonlinear modes in the open universe}

In Section II, we obtained the density contrast for a pressureless
open universe in the linear regime. We then commented that the
nonlinear modes can be obtained in the asymptotic limit and in the limit
of small times. In this section, we study an open universe in these
two extreme limits. Our analyses are not restricted to a dust universe.

It is a non-trivial task to define the {\sl asymptotic} domain in the open
Friedman universe. Indeed, in an asymptotic limit compatible with 
the Einstein equation and with any linear equation of state,
the scale factor increases linearly with time. 
This limit can be used, formally, to describe 
the asymptotic behaviour of the scale factor in the open universe, 
keeping in mind that, physically, the notion of large times is 
only meaningful when used 
relative to the Hubble time.
On the other hand, observations
are performed for {\sl small} times, namely for the
primordial universe, when open and closed universes behave as a flat
universe.

In this section, we discuss these
two limits of our solutions, 
{\it i.e.} $R\sim t$, and $R\sim t^{2/3}$. 
In the former case, the suppression of the growing modes implies that 
the inhomogeneities cease to grow after 
a sufficiently long time. Indeed, solving the series equation 
(\ref{recurs-open-md}) in the asymptotic limit gives \cite{maple}

\bear
{\delta R}_0 &=&  t \quad ,\\
{\delta R}_1 &=&  c_1 \cos\ln t + c_2 \sin\ln t\quad , \\
{\delta R}_2 &=&   c_2 \cos\sqrt 2 \ln t+ c_4 \sin\sqrt 2\ln t+\nonumber\\
&&\frac{15}{272}c_2^2t^{-1}\sin\sqrt 2\ln t \cos\left(2-\sqrt 2 \right)\ln t
-\frac{25}{272}c_1^2t^{-1}\cos\sqrt 2\ln t \cos\left(2+\sqrt 2 \right)\ln t
\nonumber\\
&&-\frac7{68}c_1c_2 t^{-1}\left(
\sin\sqrt 2\ln t  \cos\left(2-\sqrt 2 \right)\ln t
-\cos\sqrt 2\ln t \cos\left(2+\sqrt 2 \right)\ln t\right)\nonumber\\
&&+{\rm few\quad similar\quad terms}.\\
\eear
Substituting the above solutions in the time-time component 
of the Einstein equation (\ref{rho-ee}),
we recover the mass 
distribution and subsequently the following
density contrast:
\bear
\delta_1 &=& 2 c_1 \cos\ln t  +2 c_2 \sin\ln t  -6 c_1 \sin\ln t
+6c_2 \cos\ln t\quad ,\\
\delta_2 &=& -6 \sqrt 2 c_3 \sin\sqrt 2\ln t +6 \sqrt 2 c_4\cos\sqrt 2 \ln t
+{\rm few\quad similar\quad terms}.
\eear

The terms $\cos\ln t$ and  $\sin\ln t$ do not represent any growing
modes and only a decaying mode of $1/t$ is indicated by the above solutions. 
In the radiation-dominated era, however, we do find a growing
mode. In this era, the first few corrections to the scale factor 
are
\bear
{\delta R}_0 &=&  t\quad , \\
{\delta R}_1 &=& \frac 1{\sqrt t} c_1 \sin\frac{\sqrt 7}2\ln t +  
\frac 1{\sqrt t} c_2 \cos\frac{\sqrt {39}}6\ln t \quad ,\\
{\delta R}_2 &=& -\frac 1{\sqrt t} c_3 \cos\frac{\sqrt 7}2\ln t +  
\frac 1{\sqrt t} c_4 \sin\frac{\sqrt 7}2\ln t +{\cal O}\left( t^{-3}\right)
\quad ,
\eear
and consequently to the density contrast are
\bear
\delta_1 &=&-\lbrack c_1 +\sqrt{39}c_2\rbrack
\sqrt t \sin\lbrack\frac{\sqrt {39}}6\ln t\rbrack+\lbrack -c_2 +c_1 
\sqrt t\sqrt{39}\rbrack
\cos\lbrack\frac{\sqrt {39}}6\ln t\rbrack\quad , \\
\delta_2 &=&-\frac{\sqrt 7}{14}f_3(t)\sin\lbrack\frac{\sqrt 7}2 \ln t\rbrack
-3c_3\sqrt t \cos\lbrack \frac{\sqrt 7}2 \ln t\rbrack -3 c_4\sqrt t \sin 
\lbrack\frac{\sqrt 7}2 \ln t\rbrack + \frac{\sqrt 7}{14}f_4(t)
\cos\lbrack\frac{\sqrt 7}2 \ln t\rbrack
\nonumber\\
&&-3c_1c_2  \sin\lbrack\frac{\sqrt 39}6\ln t\rbrack  \cos\lbrack
\frac{\sqrt 39}6\ln t\rbrack
+\frac {7+13}4 \frac 1t c_2^2\cos^2\lbrack\frac{\sqrt 39}6\ln t\rbrack
+\frac {7+13}4 \frac 1t c_2^2\sin^2\lbrack\frac{\sqrt 39}6\ln t
\rbrack\nonumber\\
&&+\frac 52 \sqrt {39}\frac 1t (c_2^2-c_1^2) \cos
\lbrack\frac{\sqrt {39}}6\ln t\rbrack
\sin\lbrack\frac{\sqrt {39}}6\ln t\rbrack\nonumber\\
&&+\frac 12 \sqrt t f_4(t) \sin\lbrack\frac{\sqrt 7}2\ln t \rbrack
+\frac 12 \sqrt t f_3(t) \cos\lbrack\frac{\sqrt 7}2\ln t\rbrack\nonumber\\
&&-3 \sqrt{7 t}\left( c_3 \sin\frac{\sqrt 7}2\ln t-c_4  
\cos\lbrack\frac{\sqrt 7}2\ln t\rbrack \right)
\eear
where$f_3$ and $f_4$ are functions depending on integrals of
the functions already appearing above. That is,
\bear
f_3 &=& \int 25 t^{-5/2} c_1^2 \cos\left(\frac{\sqrt 2}2\ln t\right)
-\frac{38}{t^{5/2}} c_1^2\cos\left(\frac{\sqrt 2}2\ln t\right)
\cos^2\left(\frac{\sqrt {39}}6\ln t\right) +\cdots\quad ,\\
f_4 &=& \int 25 t^{-5/2} c_1^2 \sin\left(\frac{\sqrt 2}2\ln t\right)
-\frac{38}{t^{5/2}} c_1^2\sin\left(\frac{\sqrt 2}2\ln t\right)
\cos^2\left(\frac{\sqrt {39}}6\ln t\right) +\cdots\quad .
\eear
We find the growing mode $\delta\sim\sqrt t$ already at the
first order. Thus, asymptotically, the inhomogeneities can grow in an
open universe during radiation-dominated era but stop growing
during the matter-dominated era.
We interpret these results as a freezing of the present mass distribution
\footnote{ Interestingly, the seizure of the inhomogeneities
to grow after a long time 
is compatible with the numerical results of 
reference \cite{aakm} for a self-similar Newtonian universe.}.

At small times, the parametric solutions for the Friedmann scale
factor in the matter-dominated open universe approach the solution for 
the flat universe, {\it i.e.} $R\sim t^{2/3}$.
However, in the above 
perturbation scheme we cannot simply abandon the constant $k$ and approach
the result from the point of view of 
the Einstein equations for a flat universe,
since there are important contributions from the space-dependent derivatives
through $k$ ({\it e.g.} in equation (\ref{r-equation})). 
Using again a MapleV-5 program \cite{maple} we find
\bear
{\delta R}_0 &=&  c_0 t^{\frac 23}\quad , \\
{\delta R}_1 &=&  c_1 t^{\frac 23} + \cdots\quad ,\\
{\delta R}_2 &=&  c_2 t^{\frac 23} +\cdots\quad ,\\
{\delta R}_3 &=&  c_3 t^{\frac 43} +\cdots\quad .
\eear
Substituting these in the time-time component 
of the Einstein equation (\ref{rho-ee}),
we recover the mass 
distribution and subsequently the following perturbative contributions
to the density contrast:
\bear
\delta_0 &=& C_n t^{\frac 23}\quad ,\\
\delta_1 &=& C_ot^{-1}+C_p t^{\frac 23}+ \cdots\quad , \\
\delta_2 &=& C_q t^{-1}+C_s t^{-2}+C_u t^{\frac 23}+ \cdots\quad , \\
\delta_3 &=& C_v t^{-1} +C_w t^{-2}+C_x t^{-3}+ C_yt^{\frac 23}
+C_z t^{-\frac 13}+C_{aa}t^{2}+ \cdots\quad,
\eear
where $C$'s are the constant coefficients. 
Two points are worth mentioning. In the open universe, growing modes appear
earlier due to the $r$ factors in the time-time component of 
the Einstein equation (\ref{rho-ee}).
In addition to
the usual growing mode, namely $t^{\frac 23}$,
there exist new growing modes which appear at 
lower orders as compared to the flat
universe (see equation (\ref{flatdensitycontrast})). 

Moreover, later in the
expansion of the scale factor the appearance 
of new growing modes due to the inhomogeneous
part of the Einstein equation as written in (\ref{recurs-open-md}) for the
matter-dominated era proceeds as in the flat universe, but now with
enhanced consequences for the density contrast (namely they appear at
lower orders).
This profile could not be studied directly in the parametric 
solution for the matter-dominated open universe due to 
the complications arising from 
the inhomogeneous part of the 
differential equation (\ref{recurs-open-md}).

In addition to the usual growing modes, the density
contrast has an infinite number of higher-order growing modes.
These are not expected in the usual linear perturbative schemes 
\cite{weinberg,kolb,prad,coles}.
The higher growing modes were overlooked in the exact 
solutions for the open universe (\ref{diff-eq-open-sol})
because only the homogeneous part 
of the differential equation was considered, 
while the new growing modes seem to be due to a
back reaction of the system to the perturbation itself. 

The modes are divided in families, once they appear at some order of the 
perturbation, they continue appearing at higher orders together with their 
{\sl descendants} which are one order smaller in time. 

\section{Small-distance perturbation}
\vskip -.3cm
In the preceding sections, we have considered the perturbative
expansion appropriate for large distances. We now
analysis the perturbative expansion in positive powers of $r$ which is more
appropriate for small distances and thus more relevant to the
formation of galaxies and other smaller structures.
We start with the expansion
\be
R_p(t,r) =R(t)+ \sum_{n=0}^\infty {\delta R}_n(t)r^{n}\quad ,
\ee
which is substituted into the Einstein equation (\ref{r-equation}).
At order $r^{-2}$, the equation is trivially satisfied and 
the order $r^{-1}$ terms imply that
${\delta R}_1$ vanishes. The next term, however, 
involves both $R$ and ${\delta R}_2$. If we insist on 
using the homogeneous result for $R$ then the whole series disappears.
That is to say ${\delta R}_n=0$ for $n\not=0$. This problem can be understood 
once one realizes that there exists the simple separable solution
\be
R_p(t,r) =\frac{ R(t)}{1+g(t)r^2}\quad ,
\ee
which implies that  ${\delta R}_2$ is non-vanishing and proportional to $R$, 
namely ${\delta R}_2= -R g(t)$. However, we cannot find the full solution 
perturbatively.
A further 
exact solution can also be found and reads
\be
R_p(t,r) =\frac{ R(t)}r\quad .
\ee
In this case, the equation for $R(t)$ depends on the equation of state.
Furthermore, imposing the condition that
$p_r=0$ we recover the scale factor of the flat universe, namely
$R(t) \sim t^{\frac 23}$, while requiring $p_\theta=0$ we recover 
the result of the closed 
universe, namely $R(t) \sim \theta-{\rm sin}\theta$.

The structure of the small-time perturbative expansion 
in the open universe is such that new terms corresponding 
to new growing modes are
introduced into the recursive relations. For the flat 
universe we clearly find a dominant term in $t$, of the type ${\delta R}_N(t) 
\approx \frac 1N\frac {t^{\frac 23 (N +1)}}{r^{2N+1}}$ for large values
of $N$, foreseeing a logarithmic correction to the scale factor.
Moreover, the correction term itself grows at relatively small
values of time. Since $R(t) = At^{\frac 23}$, 
the Hubble constant is proportional to $A^{3/2}$, 
and we obtain ${\delta R}_N\sim A^{N+1}
\frac {t^{\frac 23 (N +1)}}{r^{2N+1}}$, and 
$\delta_N \sim A^{N}\frac {t^{\frac 23 N}}
{r^{2N+1}}$. The higher-order terms are, therefore, 
important for $t^{\frac 23}\sim
A r^2\sim H_0 r^3$, in the case of the flat universe.

\section{Conclusions}
\vskip -.3cm
\indent We analyzed the perturbation of a homogeneous 
Friedmann-Robertson-Walker
space-time as a power series in the inverse of the radial parameter
for the S-wave modes.
The perturbative solutions
depend not only on whether the universe is open, closed or flat, but also
on the epoch when the perturbation starts, namely if it starts at early
or at late times compared to the age of the universe. The standard
expressions for the density contrast in both linear 
and non-linear regimes are
contained in our results.

For the open and 
closed universes the full expression for the density contrast cannot
be obtained, since one needs to solve 
a set of inhomogeneous recursive differential 
equations for the perturbation functions, where at a given order
the lower-order solutions
become the inhomogeneous part of the differential equation. 
However, the homogeneous part of the equations
can be solved in closed form in all
cases, and contains, in addition to the usual 
modes, infinitely many growing modes.
The inhomogeneous part of the equation contains further 
valuable informations as far as the nonlinear modes are concerned. 
We have solved the full inhomogenous equation for the open
matter-dominated universe in two limits: small and large
times. At small times, the open universe 
behaves as a flat universe, with the difference that the
nonlinear growing modes appear at lower orders 
in the perturbation series.
The analysis of the large-time solutions, which covers both radiation
and matter-dominated eras, indicates 
the disappearance of the growing modes at future times.  

For the flat universe, the differential equation
simplifies since the scale factor depends explicitly on time and not
parametrically through trigonometric expressions as in the closed and
open universes. We have solved the inhomogeneous differential equation
for the flat universe and have obtained the full expression for the
linear and nonlinear growing modes (which fully agrees with the
standard results).
These modes grow faster as one goes to higher orders 
in the perturbation series. This points towards a bottom-up
scenario for the structure formation and a universe dominated by cold
dark matters.

In order to verify the observational consequences of the modes growing
faster than the standard modes of the linear regime, 
it is necessary to rewrite the results in terms of observable quantities
such as the luminosity distance
rather than the variables $(t,r)$. This is presently under
investigation \cite{amr}.

We thank F.
Brandt for his help with the Maple programs.
R.M. thanks A. Albrecht for helpful discussions.
This work was supported by Conselho Federal 
de Desenvolvimento Cient\'\i fico e
Tecnol\'ogico (CNPq-Brazil) and Funda\c c\~ao de Amparo a Pesquisa do
Estado de S\~ao Paulo (FAPESP).


\end{document}